\documentclass[10pt, b5paper]{article} 

    \usepackage[b5paper, total={5in, 8in}]{geometry}
    \usepackage[utf8]{inputenc}

    \usepackage[T1]{fontenc}
    \usepackage{microtype}
    \usepackage{sectsty}
    
    \allsectionsfont{\sffamily\mdseries}
    \usepackage[scale=0.96]{sourcesanspro}

    \usepackage[colorlinks,linkcolor=RoyalBlue,citecolor=RoyalBlue,urlcolor=RoyalBlue]{hyperref}
    \usepackage[dvipsnames]{xcolor}
    \usepackage{amsmath,stmaryrd}
    \usepackage{amssymb}
    \providecommand{\tightlist}{%
        \setlength{\topsep}{0.0ex}\setlength{\itemsep}{0.6ex}%
        \setlength{\parskip}{0pt}\setlength{\partopsep}{0.0ex}\setlength{\parskip}{0.0ex}%
    }
    
    \def\sep{\medskip\noindent}
    \def\vid{\mathsf{vid}}
    \def\vref{\rho}
    \def\Z{\mathbb{Z}}
    \def\comm{\mathsf{Comm}}

    \title{\sf Enhanced Anonymous Credentials\\ for E-Voting Systems}
    \author{\sf Tomasz Truderung \\[.6ex] \sf\normalsize Polyas GmbH}
    \date{\small\sf\today}

\begin{document}
\maketitle



\begin{abstract}

    A simple and practical method for achieving everlasting privacy in
    e-voting systems, without relying on advanced cryptographic techniques, is
    to use anonymous voter credentials. The simplicity of this approach may,
    however, create some challenges, when combined with other security
    features, such as cast-as-intended verifiability with second device and
    second-factor authentication.

    This paper considers a simple augmentation to the anonymous credential
    mechanism, using perfectly hiding commitments to link such credentials to
    the voter identities. This solution strengthens the binding between voters
    and their credentials while preserving everlasting privacy. It ensures
    that published ballots remain unlinkable to voter identities, yet enables
    necessary consistency checks during ballot casting and ballot auditing
    (cast-as-intended) to address the challenges mentioned above.

\end{abstract}

\section{Introduction}

Ballot privacy is a fundamental requirement in electronic voting systems. It
means that no one should be able to determine how an individual voter voted.
The standard notion of ballot privacy assumes that the adversary is
computationally bounded and cannot break cryptographic primitives such as
encryption or hashing. \emph{Everlasting privacy}, on the other hand, aims to
protect ballot privacy even in the presence of a \emph{computationally
unbounded adversary} that is capable of breaking cryptographic primitives such
as encryption, hashing, and digital signatures. Given this challenging
requirement, everlasting privacy usually comes with stronger trust assumptions
(a higher number of participants which need to be assumed to be trusted).
Since these two notions of ballot privacy address different threat models and
make different trade-offs on security and trust, everlasting privacy is
typically not meant to replace the standard ballot privacy, but is best seen
as complementary to it.

A simple and practical method for achieving everlasting privacy without
relying on advanced cryptographic techniques is to use \emph{anonymous voter
credentials}. In this approach, employed for instance in Belenios
\cite{CortierGG19}, an encrypted ballot published on the digital ballot box
does not refer to the real-life identity of the voter, but instead to the
voter's anonymous public credential (where the voter is in the possession of
the corresponding private credential, used to sign the ballot). The mapping
between such anonymous credentials and individual voters --- assuming the
registration process has been carried out honestly --- is known only to the
voter themselves. Consequently, the published (encrypted) ballots contain no
link to voter identities.

This simple approach has, however, some drawbacks when combined with other
security mechanisms, as discussed below.

\paragraph{\bf Individual verifiability with second device and clash attacks.}
An issue arises when anonymous credentials are combined with cast-as-intended
verifiability methods that utilize a second device, such as the one presented
in \cite{caised-evoteid, caised-tr}. In such cast-as-intended mechanisms, the
voter uses an independent second device to audit the ballot cast by the main
voting application. To do this, the second device (i) fetches the encrypted
ballot from the voting system and --- based on the data provided by the main
device and the election system --- (ii) determines the plaintext content of
the audited ballot and displays it to the voter, who can then check that it
indeed matches their intended choice. To prevent the so-called \emph{clash
attacks} \cite{KuestersTruderungVogt-SP-2012}, it is necessary to guarantee
that the ballot fetched in step (i) is \emph{uniquely} associated to the
voter. This, however, is not possible when simple anonymous credentials are
used: even if the ballot contains the voter’s public credential, the voter has
no guarantee that the same public credential has not been assigned to other
voters as well. Consequently, multiple voters could potentially access the
same ballot during the audit.

\paragraph{\bf Second-factor authentication and ``cross-voting''.} Another
problem may arise when additional (second-factor) authentication methods are
used. In this scenario, for each voter, the voting server establishes an
additional authentication method, using some form of \emph{login credentials}
such as, for instance, login/password or an electronic-ID-based
authentication. Now, to cast a ballot, the voter must be able to both
authenticate themselves toward the voting server (using the login credentials)
and submit a correctly signed ballot (for which the private credential is
needed).

Consider the scenario where a voter authenticated as A (using login
credentials of A) casts a ballot signed with credentials of voter B. When
anonymous credentials are used as described above, the voting server, not
knowing the association between public credentials and individual voters,
is not able to detect and prevent such situations. We argue, however, that
this scenario should be prevented: only a voter who uses \emph{matching}
credentials should be allowed to cast a ballot. In fact, when the above
situation happens and voter B complains, then the examination of the election
system might confirm that B's public credential was used, even though this
voter is \emph{not} recorded as having voted (have not even logged in), which
should be considered as an inconsistency of the state of the election system.
If this happens on a larger scale (affecting, say, $n$ voters), it may raise
suspicions about the integrity of the registration process. What makes this
especially problematic is that this scenario can be easily mounted by a group
of $2n$ colluding voters, in order to undermine trust in the election process,
as it is impossible to determine who is to blame: the group of voters or the
corrupted registration process.

\paragraph{\bf Our contribution.} In this paper, we consider an augmentation of
the anonymous credential mechanism which is meant to prevent both issues
discussed above. The augmentation uses perfectly hiding commitments to the
voter’s identity, which link the public credentials to specific voters without
revealing their identity. This solution strengthens the binding between voters
and their credentials while preserving everlasting privacy, under appropriate
trust assumptions: it ensures that published ballots remain unlinkable to
voter identities even for computationally unbounded adversaries, yet enables
necessary consistency checks during ballot casting and ballot auditing. We
demonstrate that our approach prevents both clash attack and cross-voting.

The used technique is similar to the one in \cite{BalogluBMP21}, but with the
following difference: in our solution, the commitments are created at
registration time and published in the public registry, instead of being
generated on the fly by the voter's client, as it is done in
\cite{BalogluBMP21} (where the issues discussed above are not addressed).

\paragraph{\bf Structure of the paper.} Section~\ref{the-chosen-solution}
provides a description of the solution, with explicit security
guarantees detailed in Section~\ref{security}. Additionally, in
Section~\ref{secret-delivery}, we identify and address some usability
challenges related to secret delivery.



\section{The Solution}\label{the-chosen-solution}

\sep\textbf{Cryptographic setup.}
We assume the standard setup for the ElGamal-based cryptosystem, with a cyclic
group $G$ of order $q$ and two predefined, independent generators $g,h \in G$.
By $\comm(x,r)$, for $x,r \in \Z_q$, we will denote the Pedersen commitment on
$x$ with randomness $r$, that is $\comm(x,r) = g^x h^r$ (note that the
expression $g^x h^r$ is interpreted over the group $G$). When $x$ is a string,
$\comm(x,r)$ will denote the Pedersen commitment to a \emph{hash} of $x$, that
is $\comm(x, r) = g^{\tilde x} h^r$, where $\tilde x = H(x) \in \Z_q$ (where
$H$ is a predefined hash function mapping strings to elements of $\Z_q$).

In the presented solution, we do not make any specific assumptions about how
ballots are encoded and encrypted and how the private and public election keys
are generated and handled. We only assume that the public (ElGamal) encryption
key is known at the time of ballot casting. We will also use a signature
scheme with a set $S$ of secret (signing) keys. 

\sep\textbf{Participants.}
The following participants are relevant for the process presented below:

\begin{itemize} \tightlist

\item a set of \emph{voters} with publicly known identifiers $\vid_1, \dots,
    \vid_n$,

\item the \emph{registrar}, responsible for generation and distribution of
    voting credentials,

\item the \emph{voting server}, responsible for voters authentication and
    handling the ballot casting process.

\end{itemize}

\subsection{Credential and registry generation} 

In the registration phase, the registrar generates the secret and public voter's
credentials, for each voter $i$ with the public identifier $\vid_i$, in the
following way:
\begin{itemize} \tightlist
    \item A secret credential \(s_i\) is randomly sampled from $S$ (the set of
        secret signing keys) and the corresponding public (verification) key
        $p_i$ is computed from $s_i$, according to the used signature scheme.

    \item A secret value $t_i$ is randomly sampled from $\Z_q$ and the
        corresponding public \emph{anonymized voter's reference} $\vref_i$ is
        computed as a perfectly hiding commitment to the voter's identity,
        with $t_i$ serving as the random commitment coin, i.e.\ 
        \begin{equation}\label{eq:rho}
            \vref_i = \comm(\vid_i, t_i)
        \end{equation}
        We will refer to $t_i$ as the voter's \emph{reference opening}.

\end{itemize}
The secret values $s_i$ and $t_i$ are securely delivered to the voter (using a
confidential channel). Once delivered, these values should be securely deleted
by the registrar\footnote{If re-delivery is needed for pragmatic reasons,
these secret credentials may need to be kept by the Registrar for some time.
In both cases, the Registrar is trusted to handle them as required by a
publicly known policy.}.  When the registration phase is over, the
\emph{public registry records} $(p_i, \vref_i)$ are published in a randomized
order (or sorted) on the public registry board.  Note that the published
records contain public credentials and public anonymized voters' references
but \emph{not} voters' identifiers. 

The idea behind this construction is that $p_i$ will serve as the public
verification key, used to check that the ballot is correctly signed by an
eligible voter (eligibility verifiability), while the anonymized voter's
reference $\vref_i$ will serve as the unique voter's identifier attached to
the encrypted ballot. On the one hand, no party will be able to link such a
ballot to the individual voter, unless this party knows the corresponding
secret $t_i$ (which is, however, intended to be known only to the voter). The
voter, on the other hand, knowing $t_i$, will be able to check and
\emph{prove}, when necessary, that the given ballot is linked to their unique
identifier $\vid_i$. This ability will play an important role, as one can see
below, at two points: at the ballot cast step and during the cast-as-intended
verification.

Independently, voters may also receive separate login credentials from the
voting server.  In this case, the knowledge of the voter's secret
credentials $s_i, t_i$ does not suffice to cast a ballot, as the
additional (matching) login credentials are needed. This provides an
additional level of protection against \emph{ballot stuffing}, in case the
voter's credentials are leaked (due to, for instance, flawed or malicious
registrar or delivery channel).

\subsection{Ballot casting}\label{sect:ballot-casting}

A voter with identifier $\vid_i$ starts the ballot casting process by entering
their secret voting credentials $s_i$, $t_i$, along with the additional login
credentials, to the voter client application. The login credentials are used
to authenticate the voter towards the voting server. The voter then selects
their choice and the client application encrypts it, following the underlying
ballot encryption method (as already noted, we do not rely on the details of
this step). Let $c_i$ denote the encrypted voter's choice.  Voter's ballot
$b_i$ is created as 
\begin{equation}\label{eq:ballot}
    b_i = (c_i, \vref_i, \sigma_i),
\end{equation}
where $\vref_i$ is the anonymized voter's reference, computed by the voter
client as $\vref_i = \comm(\vid_i, t_i)$, and $\sigma_i$ is the signature
on $(c_i, \vref_i)$ under the secret key $s_i$.

The voter client application submits the ballot $b_i$ together with $t_i$ to
the voting server.  This allows the voting server to check the consistency of
the reference $\vref_i$ included in the ballot with the corresponding public
credential and the voter's identity as follows. 

\begin{itemize}

    \item First, the voting server uses the provided value $t_i$ and the
        voter's identity $\vid_i$ established by the authentication process to
        check that value $\vref_i$ included in the ballot is in fact
        equal to $\comm(\vid_i, t_i)$. This check makes sure that the ballot is
        uniquely linked, via the reference $\vref_i$, to the identifier of the
        authorised voter. 

    \item Second, the voting server takes the public credential $p_i$
        corresponding to the voter's reference $\vref_i$, as determined by the
        content of the public registry (that is, it takes $p_i$ from the
        public registry record $(p_i, \vref_i)$) and checks that $\sigma_i$ is
        a valid signature on $(c_i, \vref_i)$ with respect to the verification
        key $p_i$.

\end{itemize}
The ballot $b_i$ is published (in the digital ballot box) only if the above
tests succeed and is rejected otherwise.  Importantly, the voting server uses
the reference opening $t_i$ only during the ballot cast time (for the above
validation steps) and discards it immediately afterwards, without storing it,
logging, or publishing it in any way.

\subsection{Ballot audit (cast-as-intended)} 

The proposed above ballot casting procedure can be combined with the
cast-as-intended method presented in \cite{caised-evoteid, caised-tr}, where
voters can use a second device to audit their ballots. In this method, the
ballot audit is enabled by the voters client application (the first device)
which displays a QR-code containing ballot audit data right after the ballot
is successfully cast.  The process consists of the following steps.

\begin{itemize}\tightlist

    \item The voter opens the ballot audit application (the second device
        application) and scans the QR-code.

    \item The second device application authenticates the voter to the
        election system (the details of the authentication process are not
        relevant here) and receives from the election system the encrypted
        ballot of the voter, together with signed acknowledgement on this
        ballot.

    \item The ballot audit application engages in an exchange of messages with
        the election system. Based on the results of this exchange and the
        data included in the QR-code, the ballot audit application determines
        the plaintext content of the ballot and displays it to the voter who
        makes sure that it indeed matches their intended choice.

    \item The signed acknowledgement of the election system is given to the
        voter as a receipt of the cast ballot.

\end{itemize}
Note that the ballot returned by the election system (which is of the form
given by \eqref{eq:ballot}) does not include the voter's identity but only the
anonymous reference $\vref_i$. To prevent so-called \emph{clash attacks}
\cite{KuestersTruderungVogt-SP-2012} (where two or more voters are lead to
audit \emph{the same ballot}), it is crucial to make sure that the returned
ballot is uniquely linked to the voter's identity.  To enable this, the first
voter's application includes the reference opening $t_i$ in the QR-code. With
this value, the second device application checks the consistency of the
reference $\vref_i$ included in the audited ballot with the voter's identity
(displayed to the voter and known to the voter up-front), by checking equation
\eqref{eq:rho}. This check eliminates the possibility that another voter was
assigned the same reference and is auditing the same ballot.

As in the case of the election system, the reference opening $t_i$ should be
used only for the above step and discarded immediately afterwards.



\section{Security Guarantees}\label{security}

Anonymous credentials, including the original, simple variant and the variant
described in this paper, do not influence standard ballot privacy (which
relies on the ballot encryption) or the universal verifiability. Similarly, the
guarantees for eligibility verifiability (the possibility to universally
verify that each ballot has been cast by an eligible voter) are not changed:
they are based on the assumption that private credentials are securely delivered
to the voters. We focus, therefore, on everlasting privacy and the prevention
of clash attacks (which is relevant for cast-as-intended).

\paragraph{\bf Everlasting privacy.}
We define everlasting privacy as ballot secrecy in the presence of
computationally unbounded adversary, that is an adversary who is not
restricted to only run polynomially bounded algorithms.  Everlasting privacy
means that even such an unbounded adversary is not able to determine choices
of individual voters, beyond the information explicitly revealed by the
election outcome.

Our solution relies on the following trust assumptions to guarantee
everlasting privacy:

\begin{itemize}\tightlist

    \item the registrar behaves honestly and follows the protocol during the
        registration phase: it does not keep nor reveal the secret values
        $s_i$ and $t_i$ to any party except to the voter themselves,

    \item the voting server acts honestly during the ballot cast time: it uses
        the provided opening $t_i$ only as specified without retaining it nor
        revealing it to a third party,

    \item the voting client is trusted (similar to standard ballot privacy
        assumptions) to not reveal the voters' choice nor the opening $t_i$,

    \item the second device acts honestly, not saving nor revealing the
        opening $t_i$ to a third party.

\end{itemize}
As a result of the above trust assumptions, we can establish that once the
voter's ballot is cast and published on the bulletin board, only the voter
themselves knows the opening $t_i$ to the anonymized reference $\vref_i$
included in this ballot. Everlasting privacy follows now directly from the
perfectly hiding property of the used commitment scheme, i.e., its ability to
perfectly (that is information-theoretically) conceal the underlying value.
In fact, given the anonymized reference $\vref_i$, it is equally likely that
it is linked to each one of the voter identifiers $\vid_1, \ldots, \vid_n$,
because for each of these identifiers, there exists exactly one opening
$t^*_j$ which matches $\vref_i$ (that is $\vref_i = \comm(\vid_j; t^*_j)$).
Even though the adversary, being computationally unbounded, can compute all
these openings, he still cannot determine which one was \emph{actually} used
and corresponds to the actual voter.

\paragraph{\bf Preventing clash attacks.}
For clash attack prevention, our solution relies on the assumption that the
used commitment scheme is computationally binding.  This means that it is
computationally infeasible (i.e.\ can be only done with negligible
probability) to produce a (commitment) $\vref$ and two openings $t_i, t_j$
which open to two different voter identifiers $\vid_i, \vid_j$, i.e.\ $\vref =
\comm(\vid_i; t_i) = \comm(\vid_j; t_j)$.  Therefore, if the second device
checks that $\rho = \comm(\vid_i; t_i)$, one can be sure that this check will
fail (with overwhelming probability) for all other voter identifiers.

Note that, in contrast to the assumptions for everlasting privacy, we now
consider a different threat model: we assume that we face a computationally
bounded adversary. We make this standard assumption, because, in this case, we
model attacks which are possible to carry out during the election time (while
the unbounded adversary considered for everlasting privacy models our
projections about capabilities of future adversaries).



\section{A practical solution for secret delivery} \label{secret-delivery}

Delivering secrets $s_i$ and $t_i$ to voters may create usability issues in
some deployment scenarios. In particular, these secrets are relatively large,
which makes it cumbersome and error-prone to type them in. There are several
solutions to this problem. In this section, we describe one possible option
and discuss its security properties.

\sep
\textbf{The solution.} The registration phase is modified in the following way:

\begin{itemize} \tightlist

    \item During the credential generation step, the values $s_i$, $p_i$,
        $t_i$, and $\vref_i$ are generated as above.

    \item Additionally, for each voter \(i\), a passcode \(\tau_i\) is
        randomly generated. From this passcode, the registrar derives, 
        using appropriate key-derivation functions, the following values: 
        (i) a symmetric-encryption key $k_i$, (ii) a \emph{login password}.

    \item The values \(s_i\) and \(t_i\) are encrypted with the key \(k_i\)
        and delivered to the voting server as \(\bar s_i, \bar t_i\),
        alongside with the voter's identifier and the \emph{hashed} login
        password (the login password will be used by the voting server for
        authentication).

    \item The passcode \(\tau_i\) is securely delivered to the voter, instead
        of directly delivering \(s_i\) and \(t_i\).

    \item Optionally, the voting server provides an independent second-factor
        authentication credentials for voters' authentication.

\end{itemize}
To initiate the ballot cast process, the voter enters their passcode $\tau_i$
--- along with the additional second-factor authentication credentials, if
used --- to the voter client application. The voter's client application
derives the login password from $\tau_i$ and uses it --- possibly with the
second-factor authentication credentials --- to authenticate the voter. After
successful authentication, the voting server responds with the encrypted
values \(\bar s_i, \bar t_i\), from which the voter's client application
decrypts \(s_i\) and \(t_i\) (using the key $k_i$ derived from the provided
passcode $\tau_i$) and proceeds to the ballot creation and casting, as
described in Section~\ref{sect:ballot-casting}.

\paragraph{\bf Ergonomics.} 
Without second-factor authentication, voters only need to provide a single
secret value, the passcode, to cast their ballots. When a second factor is
used, additional steps are required depending on the authentication method;
even then, however, the need to input long values $s_i$ and $t_i$ is avoided.

\paragraph{\bf Security.}
The described modification does not have any effect on the \emph{``standard''
ballot privacy}, on \emph{cast-as-intended verifiability}, nor on
\emph{universal verifiability}.

As for \emph{everlasting privacy}, the trust assumptions essentially stay the
same: The voting server already needed to be trusted to behave honestly during
the ballot cast time. Now, we additionally require that the voting server
keeps the values $\bar s_i, \bar t_i$ confidential.

Regarding the threat for \emph{ballot stuffing}, the situation changes when no
second-factor authentication is used. In this case a single compromised party,
the registrar, would be able to cast ballots on behalf of voters, as the
passcode $\tau_i$ alone suffices to log in and submit a correctly signed
ballot. However, when a second-factor authenticator is used, the modified
scheme preserves the security level of the original protocol: to perform
ballot stuffing, both the registrar and the voting server would need to be
compromised and collude (casting a ballot requires knowledge of the
passcode together with possession of the second-factor authenticator).



\end{document}